# Using highly uniform and smooth Selenium colloids as low-loss magnetodielectric building blocks of optical metafluids

*YongDeok Cho*[1,+], *Ji-Hyeok Huh*[1,+], *Kyung Jin Park*[1], *Kwangjin Kim*[1], *Jaewon Lee*[1], and *Seungwoo Lee*[1,2,3*]

[1]SKKU Advanced Institute of Nanotechnology (SAINT), Sungkyunkwan University (SKKU), Suwon 16419, Republic of Korea; [2]School of Chemical Engineering, Sungkyunkwan University (SKKU), Suwon 16419, Republic of Korea; [3]Department of Nano Engineering, Sungkyunkwan University (SKKU), Suwon 16419, Republic of Korea

*Email: seungwoo@skku.edu
+Equally contributed to this work

**Abstract**: We systematically analyzed magnetodielectric resonances of Se colloids for the first time to exploit the possibility for use as building blocks of all-dielectric optical metafluids. By taking synergistic advantages of Se colloids, including (i) high-refractive-index at optical frequencies, (ii) unprecedented structural uniformity, and (iii) versatile access to copious quantities, the Kerker-type directional light scattering resulting from efficient coupling between strong electric and magnetic resonances were observed directly from Se colloidal suspension. Thus, the use of Se colloid as a generic magnetodielectric building block highlights an opportunity for the fluidic low-loss optical antenna, which can be processed via spin-coating and painting.

1. Introduction

Over the last decade, all-dielectric metamaterials have hold a special position among known nanophotonic systems because the strong electric and magnetic resonances can be simultaneously induced with a simply shaped, individual meta-atom (e.g., spheres and cylindrical posts).[1–26] More importantly, as depending on the dielectric Mie resonance within high-refractive-index structure, optical loss can be remarkably reduced. In stark contrast, plasmonic counterpart which has been considered as another promising ingredient of nano-optics requires the complexed or structured meta-atoms (e.g., split ring resonators or plasmonic cluster) especially for magnetic resonances at optical frequencies; furthermore, inevitably facing a significant Ohmic loss problem.[25,26] In the early stage, the lithographic definition of subwavelength-scaled, monolithic structure made of silicon (Si), germanium (Ge), and gallium arsenide (GaAs) has been an representative class of methods in optical engineer's tool set; indeed, the materialized realizations of various intriguing and low-loss properties including index-near-zero, high-refractive-index, directional scattering, Fano resonance, and magnetic mirror effect have successfully come to the fore by benefitting from recent advances in top-down lithography.[1,5,10,11,16–20,23]

This rapidly growing field of all-dielectric nano-optics could further advance its aim by expanding the range of accessible structural motifs and fabrication methods. As a representative example, the use of high-refractive-index colloidal nanosphere (NS) as a magnetodielectric building block has been viewed as a versatile, yet pivotal strategy for low-loss optical magnetism.[3,4,6,9,13–15,21,22,24] For example, the individual Si or boron (Br) NSs themselves can exhibit the Kerr-type directional scattering by strong coupling between electric and magnetic resonance modes.[4,8,9,13–15,24] More importantly, the self-assembled Si NS cluster with an ultra-small gap (less than 5 nm), which would be difficult to achieve with top-down lithography, was found to be essential for highly directional Fano resonances and electromagnetically induced transparency (EIT).[21,22,24] In this case, the strongly confined electric dipole within the gap between Si NSs should interact with the induced magnetic dipoles within each Si NS.[22] However, contrary to low-refractive-index NSs (e.g., silica and polymer), high-refractive-index NSs have proven to be significantly difficult to attain by means of versatile chemical synthesis.[14,15] Alternatively, physical synthesis via femtosecond laser ablation has promised compelling advantages in obtaining high-refractive-index NSs; proof-of-concept demonstration of low-loss magnetodielectric resonances has been reported with this method.[4,9,13,21,22,24] Nevertheless, both intrinsic irregularity of the obtainable NPs and inability to generate NPs in copious quantities remain as obstacle for a general utility of laser ablation method. Very recently, L. Shi et al, successfully synthesized the relatively uniform Si colloids by vapor decomposition combined with high temperature annealing.[14,15] However, vapor decomposition method resulted in the hydrogenation of amorphous Si (a-Si:H) rather than pure Si;[28] consequently, the a-Si:H NSs have generally 50 ~ 70 % of pure Si NS' refractive index. Of course, high temperature annealing allows for enhancing the refractive index of a-Si:H NSs (e.g., up to 75 ~ 90 % of pure Si's refractive index).[15] However, post annealing process should give rise to the volume shrinkage of a-Si:H NSs, so that the structural fidelity of self-assembled Si NS cluster or superlattice is degraded.[15] Also, this solid-state annealing process should hinder the copious generation of uniform Si NSs in a fluid phase (colloidal suspension). Thus, it is hard to generalize such strategy for immediate practical utility.

In this work, we aim to suggest that selenium (Se) colloidal nanoparticles (NPs) can alternatively serve as a truly uniform and efficient magnetodielectric build block of optical metamaterials. It is well known that highly smooth and uniform Se NSs can be massively synthesized by one-step solution reaction at room temperature.[29] More importantly, the refractive index of Se (i.e., ~ 3.0 at visible domain) is enough to induce profound optical magnetism, while simultaneously its optical loss is negligible especially over 600 nm wavelength.[30] Nevertheless, surprisingly, there is no report on the use of Se NSs as low-loss magnetodielectric building blocks. Here, we took such synergistic advantages of Se NSs to address pressing challenges of obtaining highly uniform and low-loss magnetodielectric NPs and widen our scope of optical metamaterials. In essence, the unprecedented uniformity of Se NSs which are homogeneously dispersed in water with a high-volume fraction (at least 1 %) allowed us to observe strong electric and magnetic resonances, exactly matching with theoretical predictions; thus, we found that Se NS suspension can act as all-dielectric optical metafluids. Furthermore, highly directional scattering was observed directly from this Se colloidal metafluid, providing a platform for the liquid-state, low-loss optical antenna, which can be solution-processed via spin-coating and painting.

## 2. Results and Discussion

### 2.1 Merit of Se colloid as building block of all-dielectric metamaterial

**Figure 1a** compares the refractive indices between crystalline Si, amorphous Se, and thermally annealed a-Si:H. Here, we particularly dealt with the refractive index of annealed a-Si:H, as it has very recently become a representative colloidal NP material which exhibit a strong optical magnetism.[15,25,26] According to the previous works reported by L. Shi et al., high-temperature annealing of $Si_{0.75}H_{0.25}$ and $Si_{0.60}H_{0.40}$ at 600 °C respectively increases the refractive indices up to 90 % and 75 % of pure crystalline Si.[15] As shown in **Figure 1a**, the refractive index of Se can be still high enough to induce strong electric and magnetic resonances, even if slightly less than those of annealed a-Si:H NSs. Also, Se exhibits negligible optical loss over 600 nm wavelength, while its real part of refractive index is non-disperse.[30] Thus, Se could be also one of promising material pallet toward a direct use as low-loss magnetodielectric building blocks.

More importantly, highly uniform and smooth Se NSs with exquisite control over sizes can be chemically synthesized in copious quantities by one-step, full-solution process at room temperature.[29] In this work, we used the controlled reduction of selenious acid with assistance of hydrazine at room temperature; amorphous Se colloids were synthesized by precipitations of reduced Se. The size of the synthesized Se colloids was ranged from 190 nm to 350 nm. As-synthesized Se colloids were purified by using deionized water (i.e., repetitive centrifugation and washing).

**Figure 1b** shows a representative scanning electron microscopy (SEM) image of 230 nm Se NSs together with macroscopic images of 190 nm, 230 nm, and 300 nm Se colloidal suspensions; we figured out that highly uniform Se NSs with smooth surface were indeed obtained (size distribution was less than 3 %). Evenly distributed scattering colors from differently sized Se NSs were observed in optical microscopy images (**Figure 1c-k**): a collective set of reflective dark field (DF), reflective bright field (BF), and transmissive BF optical microscopy images for 190 nm (**Figure 1c-e**), 230 nm (**Figure 1f-h**), and 300 nm Se NSs (**Figure 1i-k**) are included as representative examples.

This versatile access to unprecedented qualities and copious quantities makes Se NSs to be starkly contrasted with annealed a-Si:H NSs. In general, a-Si:H NSs colloids can be synthesized by a decomposition of $Si_3H_8$ under high-temperature and supercritical conditions,[14,15,28] which are not easy to access especially at university level. More importantly, as-synthesized a-Si:H colloids need to be annealed at high temperature (e.g., 600 C°) to increase refractive index; solid-state annealing process severely restricts the available quantities. The volume shrinkage of a-Si:H NSs is another challenge caused by solid-state annealing process, because resulting in the deformation of self-assembled colloidal structure.[15]

The unprecedented uniformity and smoothness of colloidal Se NSs can be further elucidated by UV/VIS absorption spectroscopy. As this analysis reflects the aqueous ensemble response of Mie resonance,[31] uniformity of our Se NSs can be spectrally quantified. The left panels of **Figures 2a-c** correspond to the numerically predicted Mie resonant behaviors of 190 nm, 230 nm, and 350 nm Se NS suspensions; obviously, the UV/VIS absorption spectra for each Se NS suspension exactly matched with these theoretical predictions (i.e., width and position of peak), as shown in the right panels of **Figures 2a-c**. Particularly, all hallmarks of the Mie resonances including magnetic dipole (MD), electric dipole (ED), magnetic quadrupole (MQ), and electric quadrupole (EQ) were clearly visible. The modal characteristics of MD, ED, and MQ resonances of 230 nm Se NS are summarized in **Figures 2d**. These good agreements of UV/VIS absorption spectra with numerical simulation results strongly evidenced the high uniformity and smoothness of colloidal Se NSs; providing an opportunity for using Se colloids as alternative, yet highly efficient magnetodielectric building blocks.

## 2.2 Degradation of Se NSs during solid-state optical characterizations

Probably, it is not surprising to concern whether relatively low glass transition temperature ($T_g$) of Se (~ 31 °C)[29] could cause the degradation of the structure during optical characterizations. Because high-refractive-index nanocavity such as Se NSs can be effectively heated up via strong Mie resonance, the temperature of nanocavity undergoing resonance could be increased beyond $T_g$ of Se. The absorption and scattering spectra of aqueous Se NS suspension were found to be unchanged during repetitive measurements (e.g., UV/VIS spectroscopy). It seems that water fluid can serve as a cooler. However, it was turned out that Se NSs placed onto a solid substrate (i.e., lossy glass) were degraded during reflective DF spectroscopic measurement (see **Figures 3a** and Figure S1). In our common measurement (intensity of light source was 10 mW/m$^2$), the DF scattering spectrum of 230 nm Se NSs, obtained at the first measurement (0 min in **Figures 3a**), was already deformed from the numerical prediction (i.e., **Figure 3b**). As the time for light illumination increased, this DF spectra together with DF scattering colors were further deformed; after 5 hours, spectral peak at 650 nm originating from magnetic resonances was significantly dimmed. In contrast, DF scattering spectra of gold (Au) NSs was not changed during long-time illumination of light source, as presented in **Figures 3c-d**. From these results, we concluded that the use of Se NSs as magnetodielectric building block of all-dielectric metamaterials should be limited to fluidic platform, collectively referred to as metafluids.[31,32]

## 2.3 Se NS suspension for all-dielectric optical metafluid exhibiting directional light scattering

Despite its intrinsic limitation toward solid-state nanophotonic devices, Se colloids are still attractive, because their high uniformity together with versatile access to copious quantities makes them compelling promising for use as magnetodielectric building blocks of all-dielectric optical metafluids. To exploit this possibility in more detail, we highly concentrated 230 nm Se NS in water medium and measured DF and BF scattering spectra (see **Figure 4a-c**) of this suspension. In Supporting Movie 1 and 2, we can see the distinct Brownian motions of each individual Se colloid. As with DF optical microscopy images of dried Se NSs (**Figure 1c-k**), almost same scattering colors in reflective and transmissive DF optical microscopy images were observed across all Se NSs moving in water fluid (**Figure 4a-c**). Thus, more detailed study on the magnetodielectric resonances of Se NS (e.g., directional light scattering) can be achieved by the measurement of ensemble optical response from Se colloidal suspension as follows.

In principle, the trend of BF transmission spectrum of magnetodielectric structure (**Figure 4d**) is in inverse proportion to that of total extinction cross section;[15] indeed, the measured BF transmission spectrum of 230 nm Se colloidal suspension showed the expected trend with hallmarks of MD (i.e., dip at 665 nm) and MQ (i.e., dip at 570 nm), as presented in **Figure 4e**. The insertion of cross-analyzer in the BF transmissive pathway minimizes the electric responses;[33] as such, isolating the magnetic resonances in BF transmission spectrum. As shown in **Figure 4f**, one distinct peak at 665 nm was observed after the insertion of cross-analyzer; well matching with theoretically predicted MD scattering cross section (see Figure S2). The evidence of MQ was not visible in cross-analyzer-filtered BF transmission spectrum due to its significant loss at 570 nm (see Figure 2b and Figure S2). Thereby, we can conclude that Se colloidal suspension itself can act as all-dielectric metafluid showing strong magnetism at optical frequencies.

In addition, the directivity of light scattering from 230 nm Se colloidal suspension can be quantized by comparison between DF scattering spectra in transmission and reflection modes (**Figure 4g**).[13] For Si NSs, the wavelengths of electric and magnetic resonances were relatively far from each other mainly due to high refractive index over 3.5 at the wavelength of interest (see Figure S3a).[13] Thus, the directivity of light scattering from Si NSs is not so high. In the case of our Se NSs, MD and ED can be less confined compared with those of Si NSs, as shown in Figure S3b. In other words, both MD and ED resonances become more leaky and broad. Thereby, MD and ED resonance modes can be well overlapped (see Figure S2); as such, light scattering from Se NSs can be highly directional as predicted by numerical simulations (**Figure 4h**).[24] The experimental comparison between transmissive (forward) and reflective (backward) DF scattering spectra well validated such theoretical analysis (**Figure 4i**). Numerically analyzed far-field scattering pattern shown in **Figure 4j** confirmed such experimental spectral result of directional scattering. To the best of our knowledge, it is first time to observe the directional light scattering directly from magnetodielectric colloid suspension (i.e., optical metafluidic antennas).

## 3. Conclusion

We reimagined Se colloids to explore the possibility for use as truly uniform and low-loss magnetodielectric building blocks of fluidic metamaterials at optical frequencies (optical metafluids). By benefitting from exotic properties of Se colloids, strong magnetism and Kerker-type highly directional light scattering were measured directly from Se colloidal suspension. Thus, the use of Se NSs as magnetodielectric building blocks can be generalized toward all-dielectric optical metafluids; the soft fluidity daunting feature of metafluids can facilitate the realistic translation of optical metamaterial into the immediate practical applications.

## Methods

*Synthesis of Se colloids*: All chemical reagents including selenious acid (Aldrich, 99.999%), hydrazine hydrate ($N_2H_4 \cdot H_2O$, Aldrich, 55 % $N_2H_4$), and ethylene glycol (Aldrich, anhydrous) were used as received. 5 ml hydrazine hydrate solution (0.35 M) in ethylene glycol was added to 20 ml ethylene glycol under 350 rpm stirring rate. After 10 minutes, 0.07 M $H_2SeO_3$ dissolved in ethylene glycol was added to the hydrazine hydrate solution. To control Se NSs size, we varied the ratio of $H_2SeO_3$ to hydrazine hydrate. During the reaction, the initially transparent solution gradually turned into reddish solution within ~ 15 min; the reaction was further proceeded for about 2 hours to ensure complete reduction of $H_2SeO_3$.

*Numerical simulations*: The theoretical analyses were carried out by finite element method (FEM) and finite-difference, time-domain (FDTD) simulations. Here, FEM and FDTD were supported by COMSOL Multiphysics 5.0 and CST microwave studio 2014, respectively. FEM was used mainly for the numerical calculation of extinction cross section, while FDTD was employed for the calculation of far-field scattering pattern. The refractive indices of Se and Si used in numerical simulations (**Figure 1a**) were obtained from ref. 15 and ref. 30.

*Optical measurement*: UV/VIS absorption spectroscopy (Shimadzu) was used to experimentally measure the extinction cross section of Se colloidal suspension. To measure forward and backward scattering spectra, we used our home built DF spectroscopy, where optical microscope (Nikon Eclipse series), CCD (PIXIS 400B, Princeton Instruments), and spectrometer (IsoPlane, Princeton Instruments) are integrated. NA of objective lens for the measurement of transmission and reflection DF spectra was 0.9, respectively.

*Synthesis of Au NSs*: 75 nm sized, highly uniform Au NSs were synthesized by iterative reduction of seed and further growth method.[34] First, we synthesized Au nanorods; then, the two vertices of Au nanorods were selectively etched by a controlled reduction. We repeated reductive etching and further growth of Au until the AuNSs to be used as seed becomes highly uniform and smooth. Finally, AuNS seeds were further grown into 75 nm AuNSs.

Acknowledgement

This work was supported by Samsung Research Funding Center for Samsung Electronics under Project Number SRFC-MA1402-09.

Supporting Information

Details for degraded Se colloids and modal comparison between Si and Se colloids are included in Supporting Information.

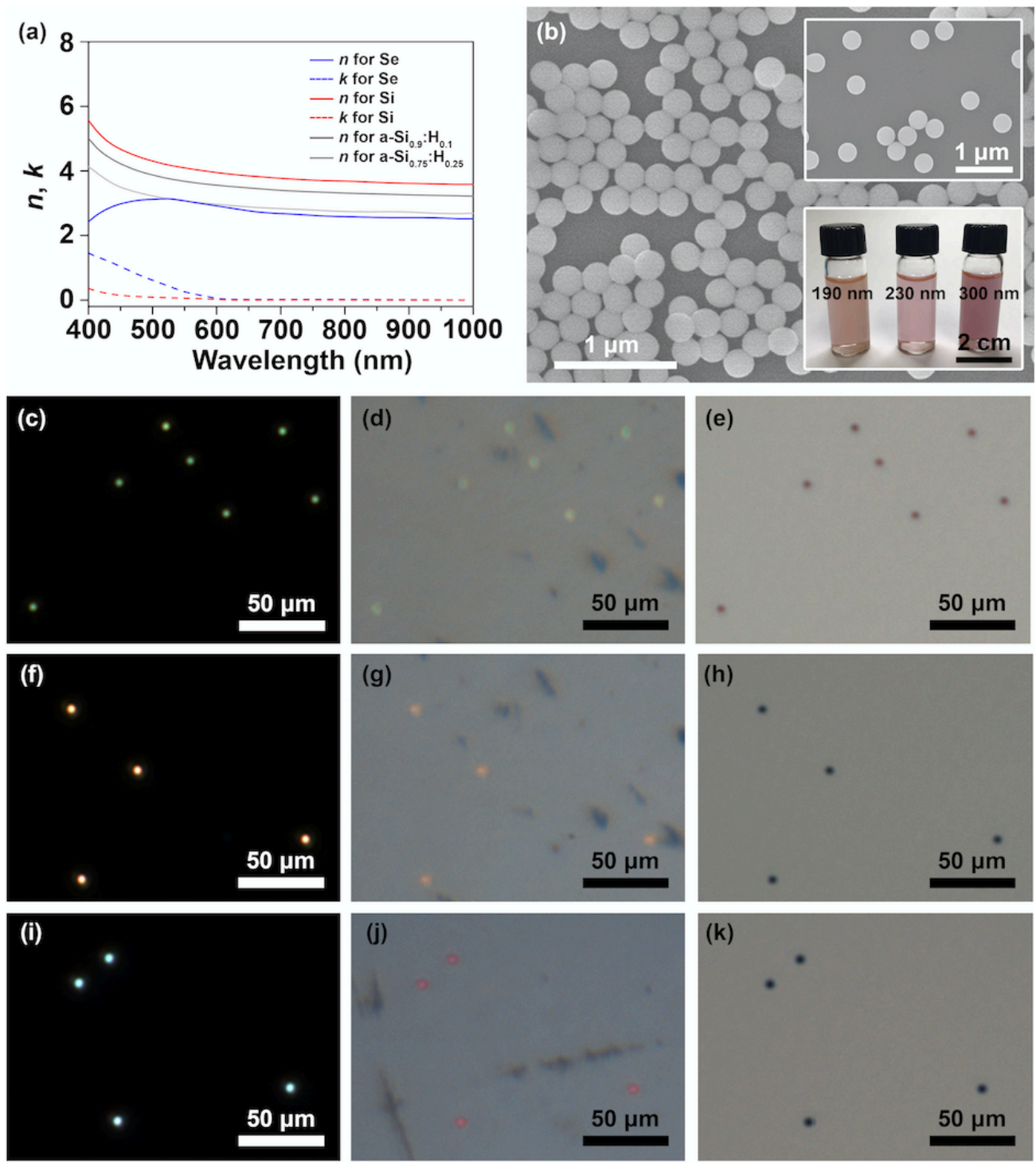

**Figure 1.** (a) Real and imaginary parts of refractive indices for crystalline Si, amorphous Se, and annealed amorphous hydrogenate Si (a-Si:H). The refractive indices of amorphous Se and annealed a-Si:H were obtained from ref. 15 and ref. 30, respectively. (b) Scanning electron microscopy (SEM) image of 230 nm Se nanospheres (NSs). The two inset images correspond to the magnified SEM image of 230 nm Se NSs and macroscopic images of 190 nm, 230 nm, and 300 nm Se colloidal suspension. (c-k) A collective set of optical microscopy images for (c-e) 190 nm, (f-h) 230 nm, and (i-k) 300 nm Se NSs. Here, Se NSs were placed onto a solid-state substrate (glass). Each row consists of reflective dark field (DF), reflective bright field (BF), and transmissive BF optical microscopy images from left to right panels.

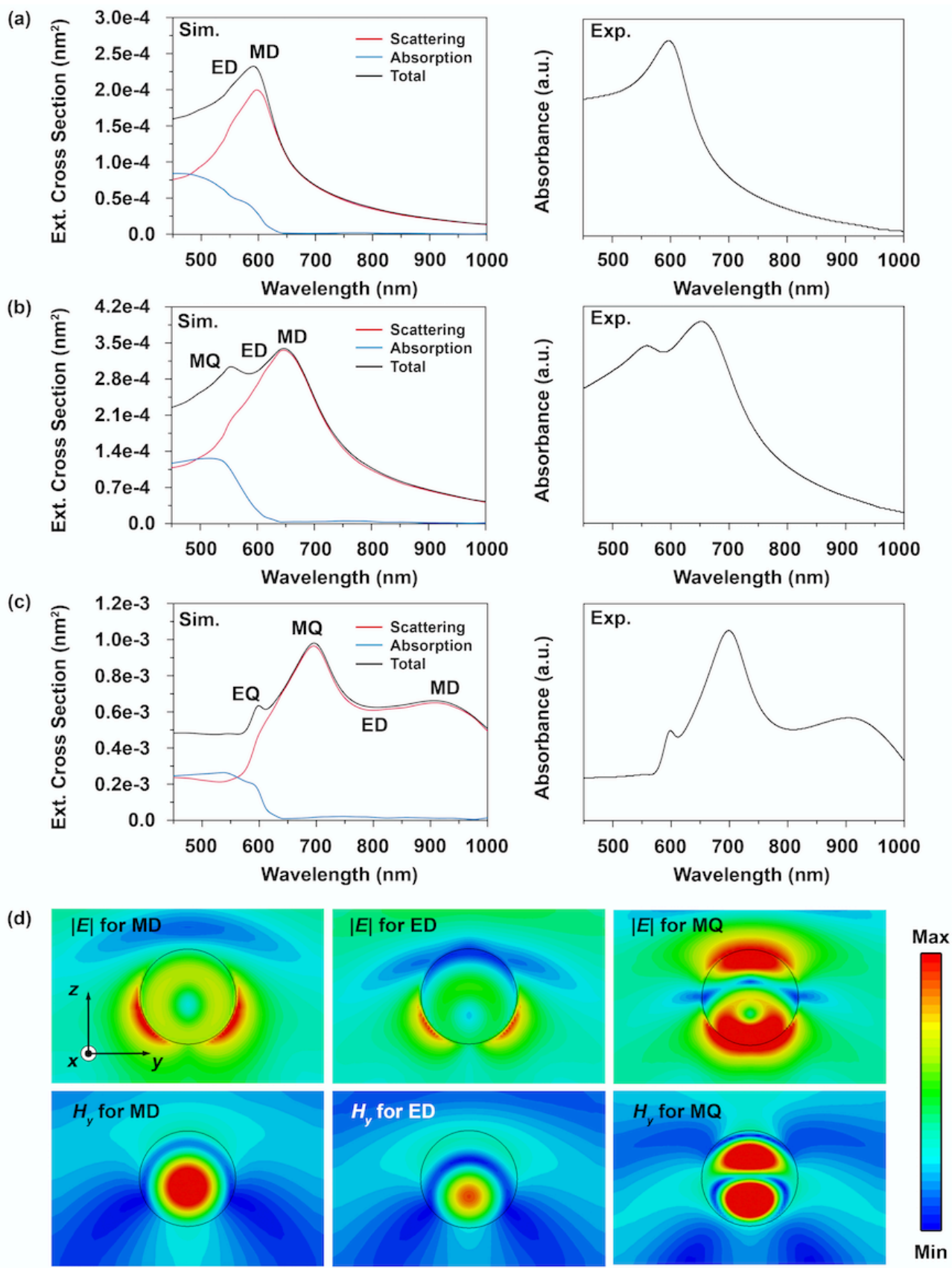

**Figure 2.** (a-c) Magnetodielectric resonances of (a) 190 nm, (b) 230 nm, and (c) 350 nm Se NSs, dispersed in water medium. In each part, left panels correspond to the theoretical analyses done by numerical simulation, while right panels indicate the results of UV/VIS absorption spectroscopy. (d) Modal analysis of magnetic dipole (MD), electric dipole (ED), and magnetic quadrupole (MQ)

for 230 nm Se NS dispersed in water. The position of ED can be found in Figure S2b. Here, light was assumed to be illuminated along *z*-axis (from top to bottom).

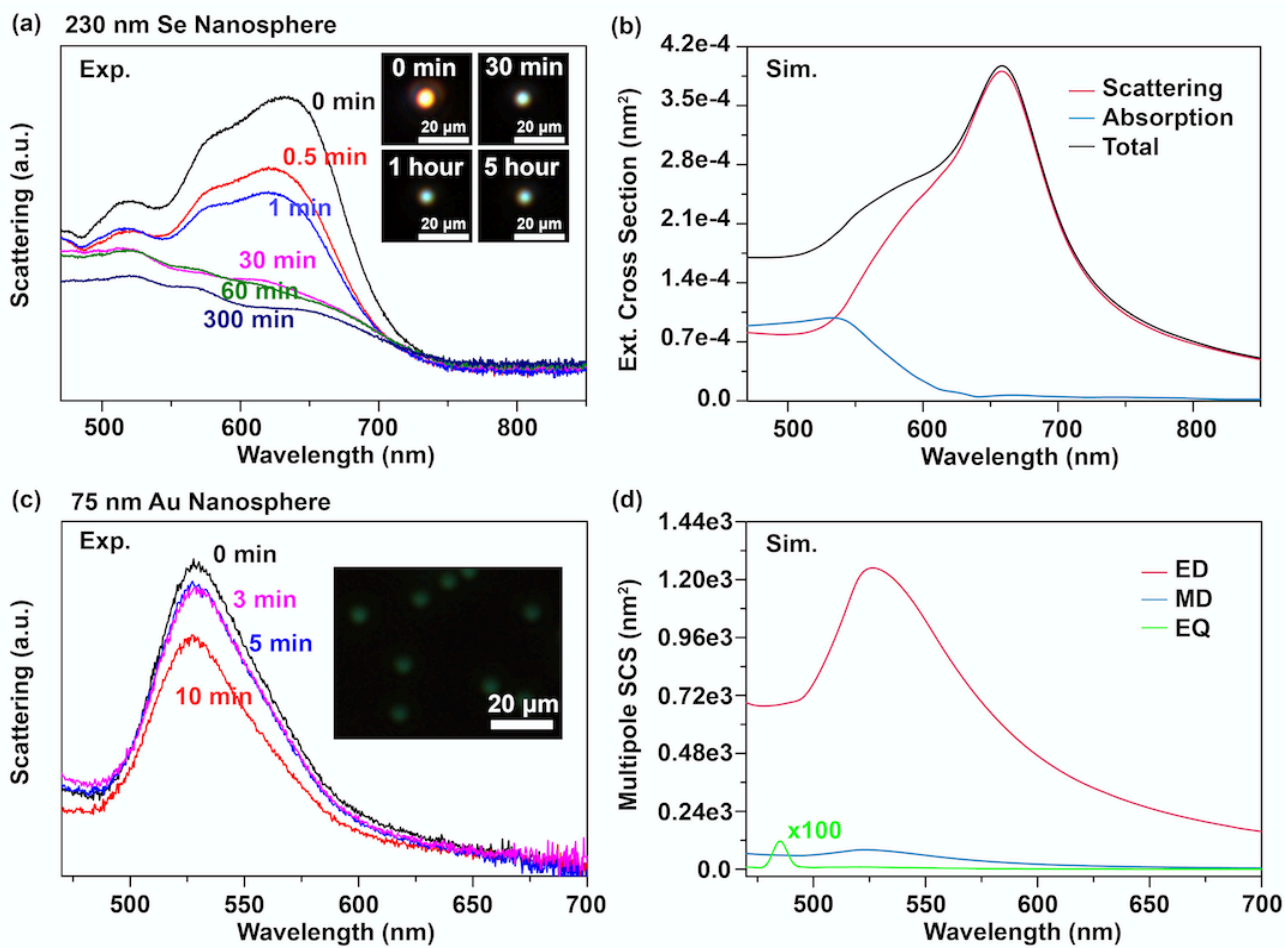

**Figure 3.** (a) DF scattering spectra of 230 nm Se NSs as function of light illumination time. Insets indicate DF optical microscopy images. (b) Theoretical analyses of DF scattering for 230 nm Se NSs placed on solid-state glass substrate. (c) DF scattering spectra of 75 nm Au NS as function of light illumination time. Inset indicates DF optical microscopy image of 75 nm Au NSs placed on solid-state glass substrate. (d) Theoretical analyses of scattering cross section (SCS) for 75 nm Au NS placed on solid-state glass substrate. The contributions of ED, MD, and EQ to scattering cross section are numerically analyzed.

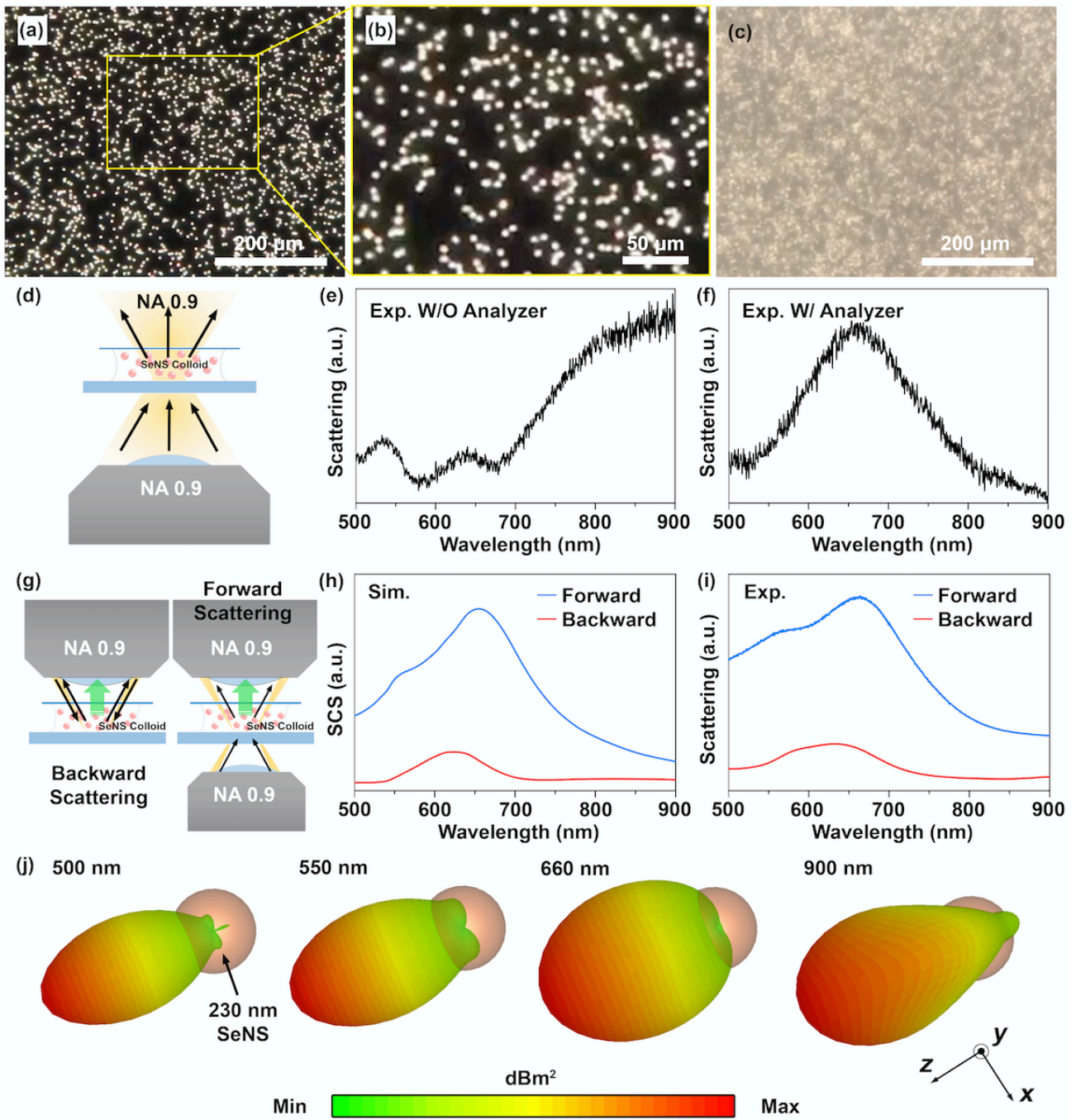

**Figure 4.** (a-b) Reflective and (c) transmissive DF images of 230 nm Se NSs dispersed in water (230 nm Se NSs metafluids). (d) Schematic for the measurement setup of BF transmission spectra. (e-f) Transmissive BF spectra of 230 nm Se NSs metafluids (e) without and (f) with cross-analyzer. (g) Schematic for the measurement setup of DF transmissive (i.e., forward scattering) and reflective (i.e., backward scattering) spectra. (h) Numerically simulated scattering cross section (SCS) and (i) experimentally measured forward/backward scattering spectra of 230 nm Se NSs metafluids. These spectra were obtained from ensemble of numerous 230 nm Se NSs dispersed in water. (j) Theoretically analyzed far-field scattering pattern as function of wavelength. Here, source light was assumed to be illuminated along *z*-axis (from right to left).

# Using highly uniform and smooth Selenium colloids as low-loss magnetodielectric building blocks of optical metafluids

*YongDeok Cho*[1,+], *Ji-Hyeok Huh*[1,+], *Kyung Jin Park*[1], *Kwangjin Kim*[1], *Jaewon Lee*[1], and *Seungwoo Lee*[1,2,3*]

[1]SKKU Advanced Institute of Nanotechnology (SAINT), Sungkyunkwan University (SKKU), Suwon 16419, Republic of Korea; [2]School of Chemical Engineering, Sungkyunkwan University (SKKU), Suwon 16419, Republic of Korea; [3]Department of Nano Engineering, Sungkyunkwan University (SKKU), Suwon 16419, Republic of Korea

*Email: seungwoo@skku.edu

[+]Equally contributed to this work

Keywords: Dielectric metamaterials, Magnetism, Colloids, Se nanoparticles, Optical metafluids

1. Degradation of Se colloids during solid-state optical measurement

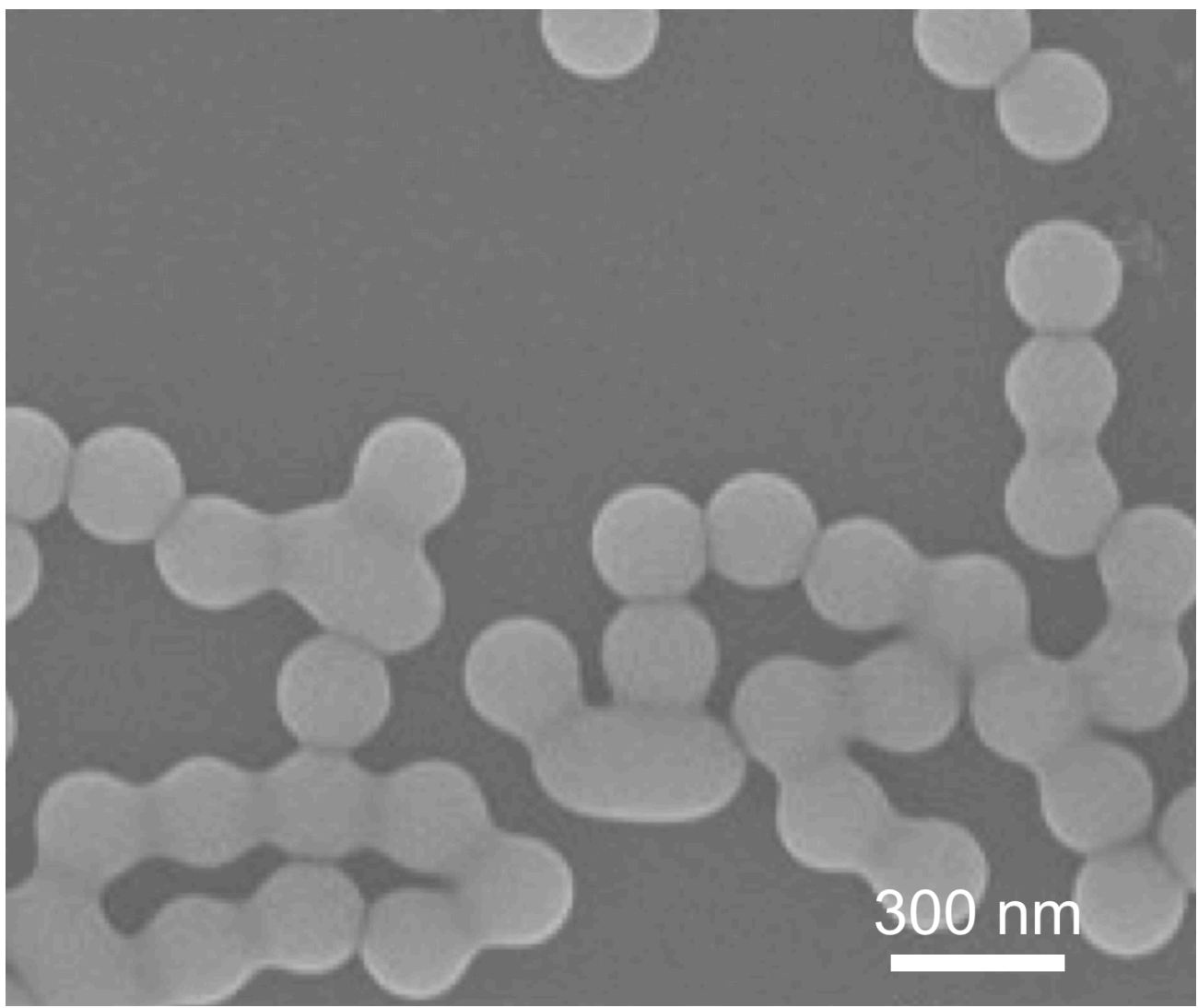

**Figure S1**. SEM image of Se colloids, which were exposed to light source of DF spectroscopy for 30 min.

## 2. Contribution of electric and magnetic resonances to light scattering of Se colloids

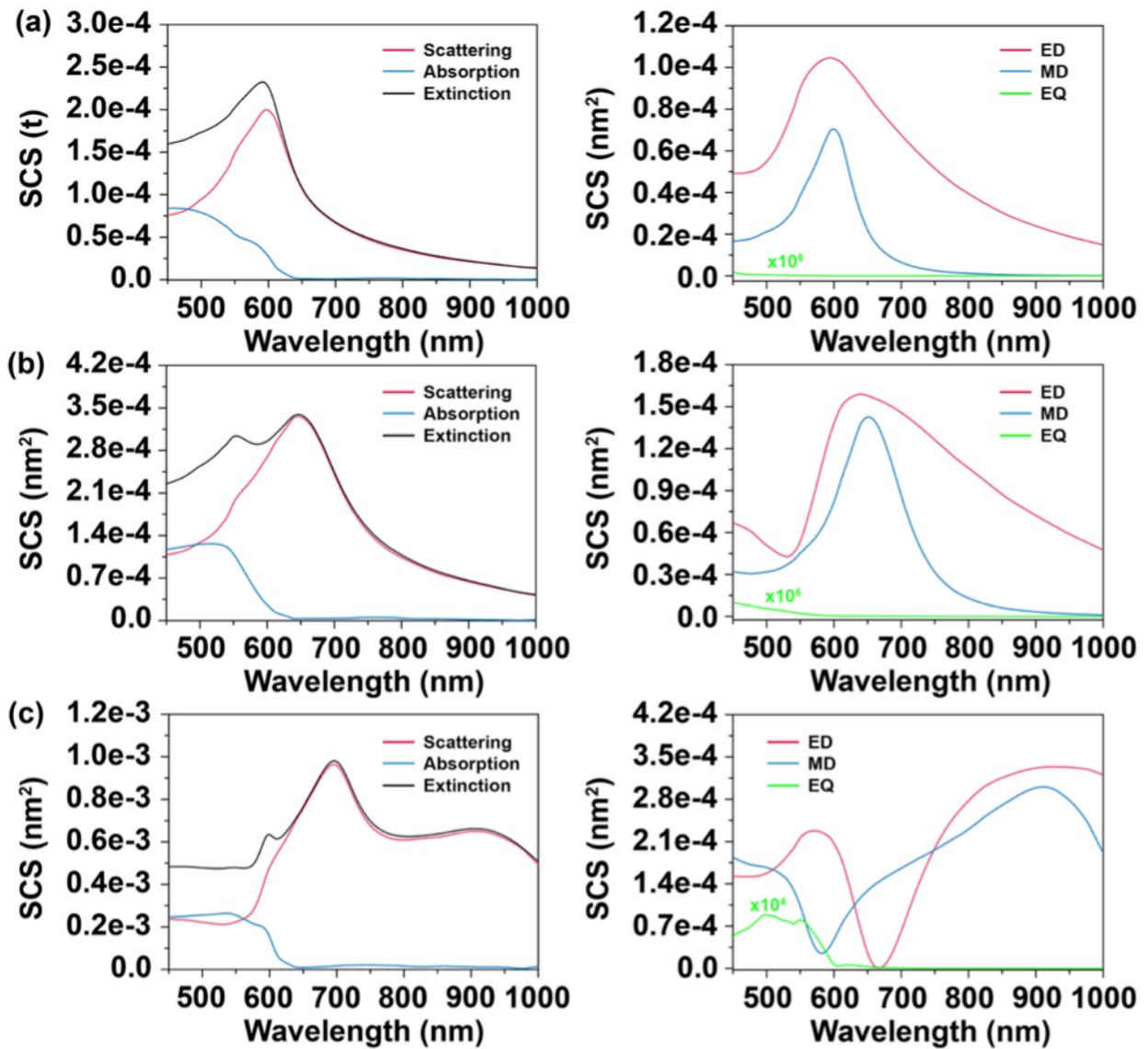

**Figure S2**. Contribution of magnetic dipole (MD), electric dipole (ED), and electric quadrupole (EQ) resonances to light scattering of Se colloids. (a) 190 nm, (b) 230 nm, and (c) 350 nm.

3. Mode comparison between 230 nm Si and Se nanospheres

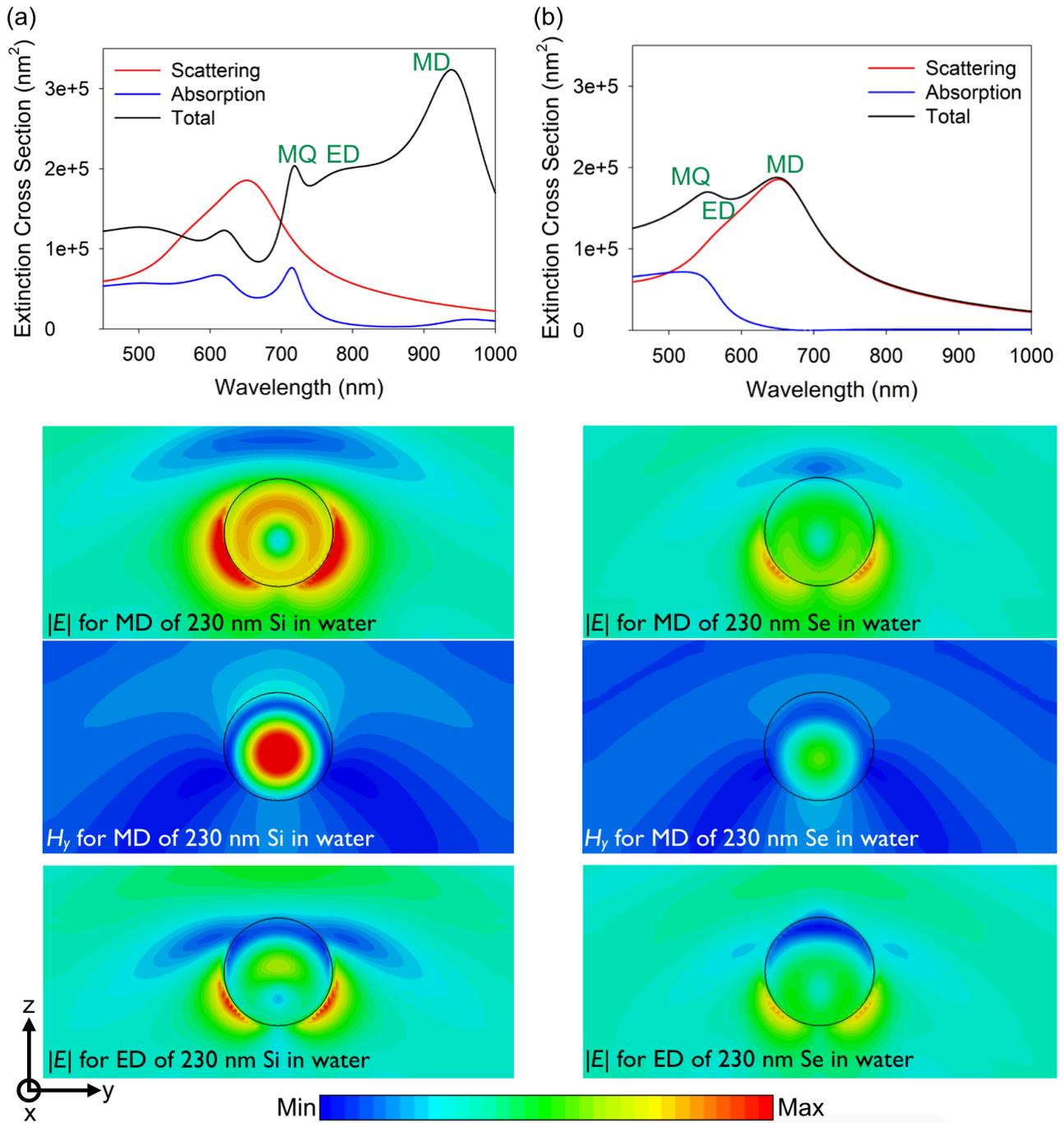

**Figure S3**. Comparisons of MD and ED resonances between (a) Si nanosphere and (b) Se nanosphere. These nanospheres have 230 nm in diameter.